\documentclass[aps,preprint]{revtex4}%
\usepackage{amsfonts}
\usepackage{amsmath}
\usepackage{amssymb}
\usepackage{graphicx}%
\usepackage{caption}
\usepackage{subcaption}
\providecommand{\U}[1]{\protect\rule{.1in}{.1in}}

\begin{document}
\title{General Relativistic Fall on a Thick-Plate}
\author{M. Halilsoy}
\email{mustafa.halilsoy@emu.edu.tr}
\author{V. Memari}
\email{vahideh.memari@emu.edu.tr}
\affiliation{Department of Physics, Faculty of Arts and Sciences, Eastern Mediterranean
University, Famagusta, North Cyprus via Mersin 10, Turkey}
\date{\today}

\begin{abstract}
As an extension of a thin-shell, we adopt a single parametric plane-symmetric
Kasner-type spacetime to represent an exact thick-plate. This naturally extends the
domain wall spacetime to a domain thick-wall case. Physical properties of such
a plate with symmetry axis $z$ and thickness $0\leq z\leq z_{0}$ are
investigated. Geodesic analysis determines the possibility of a Newtonian-like
fall, namely with constant negative acceleration as it is near the Earth's
surface. This restricts the Kasner-like exponents to a finely-tuned set, which together with the thickness and energy parameter determine the G-force of the plate. In contrast to the inverse square law, the escape velocity of the thick-shell is unbounded. The
metric is regular everywhere but expectedly the energy-momentum of the
thick-plate remains problematic.

\end{abstract}

\maketitle

\section{Introduction}

We adopt a class of plane symmetric spacetimes from which we define a metric to
represent a plate of constant thickness. The spacetime is analogous to the
Kasner cosmological model \cite{1} in which time $\left(  t\right)  $ is replaced by
the spacelike coordinate $\left(  z\right)  $ and the exponents satisfy the
Kasner conditions. Our idea is to establish a geometrical thick-shell and
analyze the fall of a particle on such a plate with thickness $z_{0}$. For a
symmetric plate, we may adopt $z<\left\vert z_{0}\right\vert $, however, due
to the symmetry we confine our plate to $0<z<z_{0}$ with $-\infty<x,y<+\infty
$. In the limiting case of $z_{0}\rightarrow0$, we naturally recover the case
of a thin-shell of Dirac delta thickness $\delta\left(  z\right)  $. Our aim
is to find a regular geometry that is free of singularities. To obtain this we
observe that the stress tensor component along $z$, namely $T_{zz}$ must
vanish. Our choice for the spacetime metric is%
\begin{equation}
ds^{2}=f^{2k\left(  k-1\right)  }\left(  dt^{2}-dz^{2}\right)  -f^{2k}%
dx^{2}-f^{2\left(  1-k\right)  }dy^{2}\label{1}%
\end{equation}
where $f=f\left(  z\right)  $ and the constant parameter $k$ is related to the
Kasner exponents. The function $f\left(  z\right)  $ has a particular
structure, which is finely tuned to reduce to the form%
\begin{equation}
f\left(  z\right)  =a+bz\label{2}%
\end{equation}
with specific constants $a$ and $b$. Effectively, our choice will be such that
$a=0$ and $b=1$, for $z>z_{0}$, which is exterior to the plate. Although not
compulsory such a choice will provide the simplest case to describe a regular
thick shell. Another reason will be to compare our spacetime with a domain
wall geometry in which $z\rightarrow\left\vert z\right\vert $. \cite{2}\cite{3}\cite{4} The geometry
extends a thin domain wall into a thick wall in a natural way. Our choice of
$f\left(  z\right)  $ \ will effectively be of the form \cite{5}
\begin{equation}
f\left(  z\right)  =\left\{
\begin{array}
[c]{c}%
z\text{, for \ \ \ \ \ \ \ \ \ \ \ \ \ \ \ \ \ \ \ \ \ \ \ \ \ \ }z>z_{0}\\
-z\text{, for \ \ \ \ \ \ \ \ \ \ \ \ \ \ \ \ \ \ \ \ \ \ \ \ \ \ }z<-z_{0}\\
\cosh a_{0}z\text{, for \ \  \ \ \ \ \ }-z_{0}<z<z_{0}%
\end{array}
\right.  \label{3}%
\end{equation}
in which $a_{0}=$constant. Due to the perfect symmetry about $z$ and for
technical simplicity, we shall consider our plate to lie in $0<z<z_{0}>0$,
without loss of generality. Through geodesics, we shall compare the Newtonian
and general relativistic falls on such a plate. The role played by the
parameter $k$ will be crucial: for $k=\frac{1}{2}$, for instance, we have
negative constant acceleration toward the plate. Not to mention, for $k=0,1$,
the exterior plate spacetime becomes flat. For $z>z_{0}$, our metric will be a
vacuum, but yet not flat for $k\neq0,1$. Our work may be considered as an extension to the problem of 'fall' to a thin plane \cite{6}. 

A drawback of our model will be that the energy conditions such as Weak, Null
and Strong will not be satisfied. In that case, our thick-shell can be
considered as an exotic one. However, by coupling additional fields, which is
not our aim in this study, the fate of our plate changes from exotic to real,
as in the case of general domain wall problems.

The organization of the paper is as follows. In section II we introduce our
plate geometry, with energy-momentum analysis, show correspondence with the
Kasner model, and analyse the second fundamental form. The general
relativistic fall through geodesic analysis will be considered in section III.
We complete the paper with our conclusion in section IV. Our notation
throughout is of metric signature $\left(  +,-,-,-\right)  $, with speed of
light $c=1$, Newtonian constant, satisfying $8\pi G=1$ and Einstein equation
$G_{\mu\nu}=-T_{\mu\nu}$.

\section{The Thick-Plate Geometry}

\begin{figure}
\centering
\begin{subfigure}[b]{0.4\textwidth}
\centering
\includegraphics[width=\textwidth]{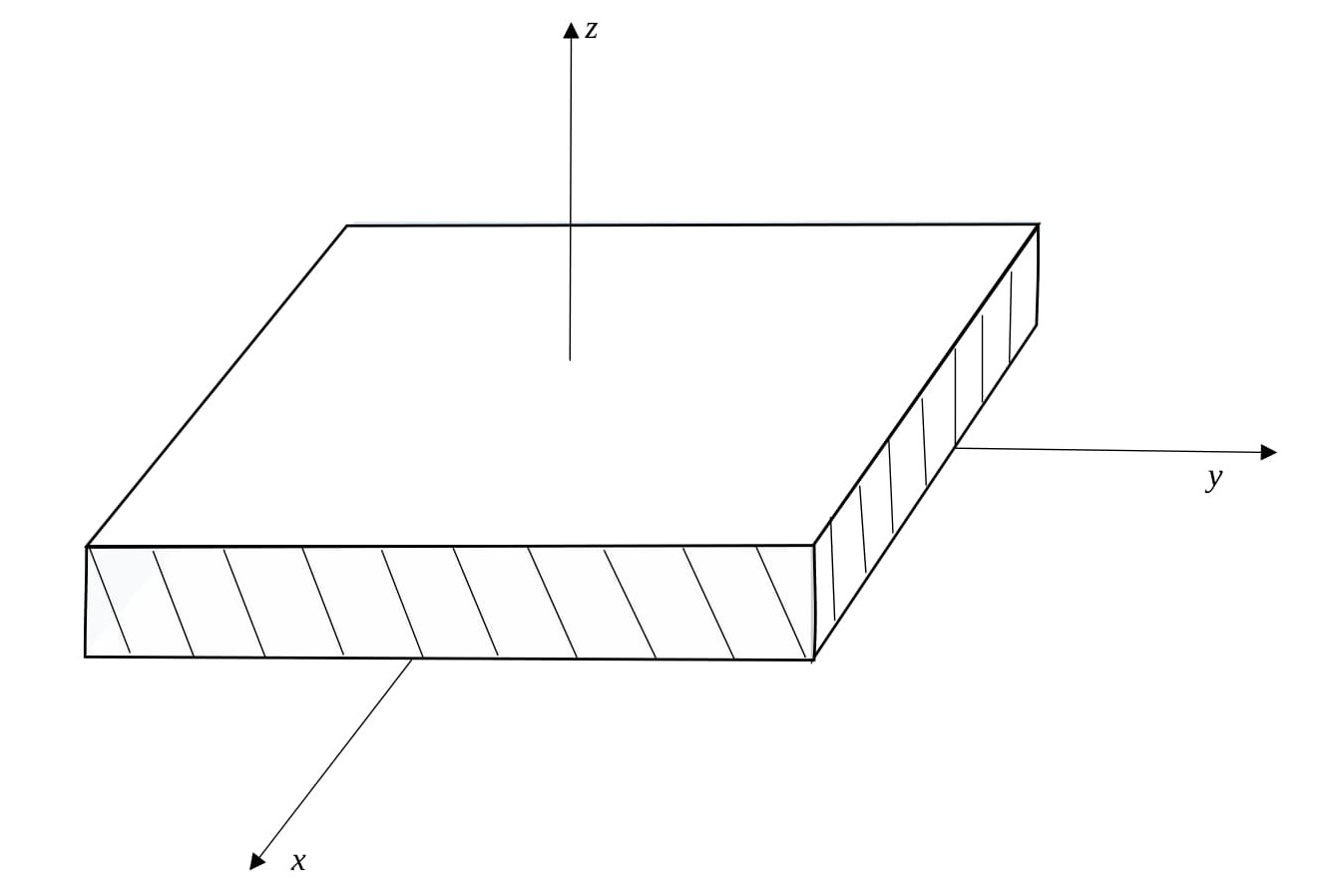}
\caption{The 3D picture of the plate.}
\label{Fig1a}
\end{subfigure}
\hfill
\begin{subfigure}[b]{0.5\textwidth}
\centering
\includegraphics[width=\textwidth]{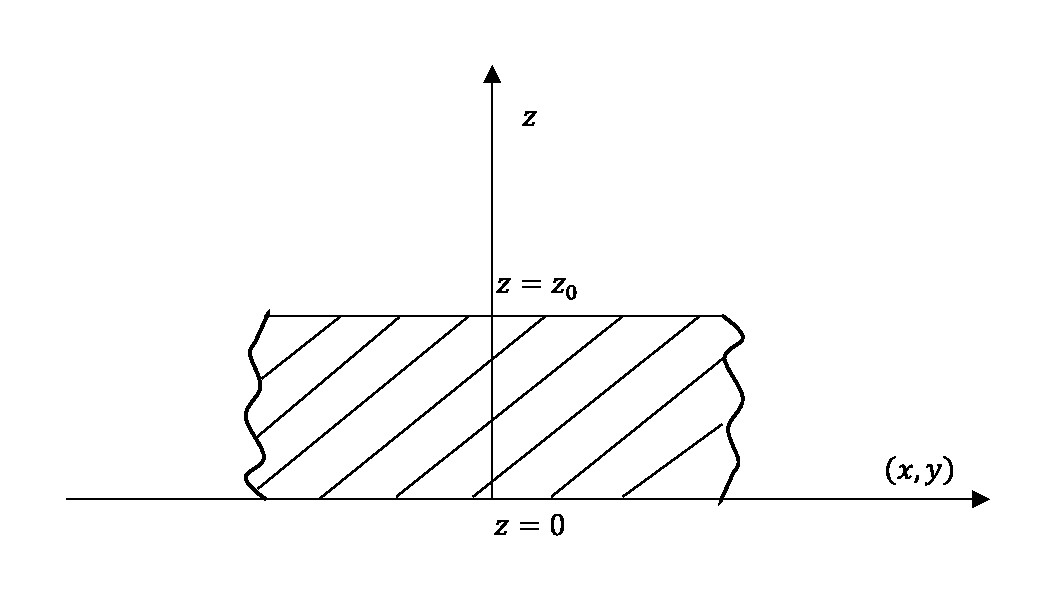}
\caption{The cross-sectional view of the plate for $0\leq z\leq z_{0}$ and $-\infty<x,y<+\infty$.}
\label{Fig1b}
\end{subfigure}
\hfill
\caption{Plot of an infinite thick-plate of thickness $z_{0}$}
\label{fig1}
\end{figure}

The line element (\ref{1}) in which the metric function $f\left(  z\right)  $
depends only on $z$ admits the obvious Killing vectors $\delta_{\mu}^{t}$,
$\delta_{\mu}^{x}$ and $\delta_{\mu}^{y}$. In the most general form, our
choice for $f\left(  z\right)  $ will be \cite{5}%
\begin{align}
f\left(  z\right)   &  =\left.  \cosh\left(  a_{0}\left[  z\left(
\Theta\left(  z\right)  +\Theta\left(  -z\right)  \right)  -\left(
z-z_{0}\right)  \Theta\left(  z-z_{0}\right)  -\left(  z+\left\vert
z_{0}\right\vert \right)  \Theta\left(  -z-\left\vert z_{0}\right\vert
\right)  \right]  \right)  \right.  +\nonumber\\
&  \left.  a_{0}\left(  \sinh a_{0}z_{0}\right)  \left(  z-z_{0}\right)
\Theta\left(  z-z_{0}\right)  -a_{0}\left(  \sinh a_{0}z_{0}\right)  \left(
z+\left\vert z_{0}\right\vert \right)  \Theta\left(  -z-\left\vert
z_{0}\right\vert \right)  \right.  \label{4}%
\end{align}
where $\left(  a_{0},z_{0}\right)  $ are constants and $\Theta\left(
z\right)  $ stands for the Heaviside step function. The function $f\left(
z\right)  $ in (\ref{4}) represents a plate confined between $-\left\vert
z_{0}\right\vert \leq z\leq z_{0}$. By the symmetry of the problem we need to
consider only the sector $z>0$ so that instead of (\ref{4}) we will employ%
\begin{equation}
f\left(  z\right)  =\cosh\left(  a_{0}\left[  z\Theta\left(  z\right)
-\left(  z-z_{0}\right)  \Theta\left(  z-z_{0}\right)  \right]  \right)
+a_{0}\left(  \sinh a_{0}z_{0}\right)  \left(  z-z_{0}\right)  \Theta\left(
z-z_{0}\right)  \label{5}%
\end{equation}
Physically $z_{0}$ will be the thickness of the shell and $a_{0}^{2}$ is a
measure of energy. The $\Theta\left(  z\right)  $ satisfies the standard
condition for a step function%
\begin{equation}
\Theta\left(  z\right)  =\left\{
\begin{array}
[c]{c}%
1\text{, \ \ \ \ \ \ \ }z>0\\
0\text{, \ \ \ \ \ \ \ }z<0
\end{array}
\right.  \label{6}%
\end{equation}
whose derivative is the Dirac delta function%
\begin{equation}
\delta\left(  z\right)  =\frac{d}{dz}\Theta\left(  z\right)  \label{7}%
\end{equation}
From the expression (\ref{5}) our plate is confined in $0<z<z_{0}$, and in
$\left(  x,y\right)  $ directions it extends throughout the $\left(
x,y\right)  $ coordinates. (FIG. \ref{fig1}).
The interesting property of the function $f\left(  z\right)  $ in (\ref{5}),
which makes the main motivation of this study is that it satisfies the
distributional property%
\begin{equation}
\frac{f^{^{\prime\prime}}\left(  z\right)  }{f\left(  z\right)  }=a_{0}%
^{2}\left(  \Theta\left(  z\right)  -\Theta\left(  z-z_{0}\right)  \right)
\label{8}%
\end{equation}
where a 'prime' means $\frac{d}{dz}$. By virtue of this condition, we have
$f^{^{\prime\prime}}\left(  z\right)  =0$, for $z>z_{0}$, and $f\left(
z\right)  $ must be of the form $f\left(  z\right)  =a+bz$. For $z<z_{0}$ on
the other hand we have $f\left(  z\right)  =\cosh\left(  a_{0}z\right)  $,
which corresponds to the inner solution. In summary, we have%
\begin{equation}
f\left(  z\right)  =\left\{
\begin{array}
[c]{c}%
a+bz\text{, \ for \ \ \ \ \ \ \ \ \ \ \ \ \ }z>z_{0}\\
\cosh a_{0}z\text{, for \ \ \ \ \ \ }0<z<z_{0}%
\end{array}
\right.  \label{9}%
\end{equation}
where%
\begin{align}
a  &  =\cosh a_{0}z_{0}-a_{0}z_{0}\sinh a_{0}z_{0}\nonumber\\
b  &  =a_{0}\sinh a_{0}z_{0} \label{10}%
\end{align}
We tune now our parameters such that $a=0$ and $b=1$. This particular choice
admits a set of numbers given by%
\begin{align}
a_{0}  &  =0.66\text{ \ \ \ , \ \ \ \ \ }z_{0}=1.81\nonumber\\
a_{0}z_{0}  &  \simeq1.99 \label{11}%
\end{align}
With this choice what we have achieved is that our spacetime takes the two
possible cases
1. The outer plate metric, for $z>z_{0}$%
\begin{equation}
ds^{2}=z^{2k\left(  k-1\right)  }\left(  dt^{2}-dz^{2}\right)  -z^{2k}%
dx^{2}-z^{2\left(  1-k\right)  }dy^{2} \label{12}%
\end{equation}

2. The inner plate metric, for $0<z\leq z_{0}$%
\begin{equation}
ds^{2}=\left(  \cosh a_{0}z\right)  ^{2k\left(  k-1\right)  }\left(
dt^{2}-dz^{2}\right)  -\left(  \cosh a_{0}z\right)  ^{2k}dx^{2}-\left(  \cosh
a_{0}z\right)  ^{2\left(  1-k\right)  }dy^{2}\label{13}%
\end{equation}
In the following sections we shall derive the Kasner connections, the
energy-momentum and investigate the extrinsic curvature of the line elements
(\ref{12}) and (\ref{13}).

\subsection{Relation with the Kasner model for $z>z_{0}$}

For the line element (\ref{12}) we introduce a new coordinate $Z\left(
z\right)  $ defined by%
\begin{equation}
dZ=z^{k\left(  k-1\right)  }\left(  dz\right)  \label{14}%
\end{equation}
Upon integration and substitutions into (\ref{12}) with appropriate scalings
$\left(  t,x,y,z\right)  \rightarrow\left(  \widetilde{t},\widetilde
{x},\widetilde{y},Z\right)  $ we cast the line element (\ref{12}) into%
\begin{equation}
ds^{2}=Z^{2p_{1}}d\widetilde{t}-dZ^{2}-Z^{2p_{2}}d\widetilde{x}^{2}-Z^{2p_{3}%
}d\widetilde{y}^{2}\label{15}%
\end{equation}
It can easily be checked that the exponents%
\begin{align}
p_{1} &  =\frac{k\left(  k-1\right)  }{k\left(  k-1\right)  +1}\nonumber\\
p_{2} &  =\frac{k}{k\left(  k-1\right)  +1}\nonumber\\
p_{3} &  =\frac{1-k}{k\left(  k-1\right)  +1}\label{16}%
\end{align}
satisfy the well-known Kasner conditions%
\begin{align}
p_{1}+p_{2}+p_{3} &  =1\nonumber\\
p_{1}^{2}+p_{2}^{2}+p_{3}^{2} &  =1\label{17}%
\end{align}
The Kretchmann scalars, $K=R_{\mu\nu\rho\sigma}R^{\mu\nu\rho\sigma}$ for
$z>z_{0}$ and for $z<z_{0}$ are given respectively by%
\begin{equation}
K=16k^{2}z^{4\left(  k-k^{2}-1\right)  }\left(  k^{4}-3k^{3}+4k^{2}%
-3k+1\right) \hspace{6cm}z>z_{0}\label{18}
\end{equation}
and
\begin{equation}
\begin{split}
K&=4a_{0}^{4}\left(\cosh{a_{0}z}\right)^{4\left(k-k^{2}-1\right)}\\
& \times\Bigl[\left(k^{2}-k+1\right)^{2}\left(\cosh{a_{0}z}\right)^{4}-6k^{2}\left(k-1\right)^{2}\left(\sinh{a_{0}z}\right)^{2}\left(\cosh{a_{0}z}\right)^{2}\\
 &+4k^{2}\left(k-1\right)^{2}\left(k^{2}-k+1\right)\left(\sinh{a_{0}z}\right)^{4}\Bigr]
 \hspace{7cm}z<z_{0}
\label{19}
\end{split}
 \end{equation}
Expectedly the scalar $K$ is everywhere finite and our spacetime is completely regular.

\subsection{Global form of the energy-momentum tensor}

From the Einstein equations $G_{\mu\nu}=-T_{\mu\nu}$, (our choice has
$\frac{8\pi G}{c^{4}}=1$) we have for the inner region%
\begin{align}
T_{tt} &  =a_{0}^{2}\left(  \Theta\left(  z\right)  -\Theta\left(
z-z_{0}\right)  \right)  \nonumber\\
T_{xx} &  =-\left(  k-1\right)  ^{2}\left(  \cosh a_{0}z\right)  ^{2k\left(
k-1\right)  }T_{tt}\nonumber\\
T_{yy} &  =-k^{2}\left(  \cosh a_{0}z\right)  ^{2\left(  1-k\right)  }%
T_{tt}\nonumber\\
T_{zz} &  =0\label{20}%
\end{align}
It is observed that for $z>z_{0}$, $T_{\mu\nu}=0$, but the spacetime is not
flat unless $k=0,1$.

The standard fluid form of the $T_{\mu\nu}$ can be expressed by%
\begin{equation}
T_{\mu}^{\nu}=diag\left(  -\rho,p_{x},p_{y},p_{z}\right)  \label{21}%
\end{equation}
which implies that%
\begin{align}
\rho &  =-a_{0}^{2}\left(  \cosh a_{0}z\right)  ^{-2k\left(  k-1\right)
}\nonumber\\
p_{x}  &  =a_{0}^{2}\left(  k-1\right)  ^{2}\left(  \cosh a_{0}z\right)
^{2k\left(  k-2\right)  }\nonumber\\
p_{y}  &  =a_{0}^{2}k^{2}\left(  \cosh a_{0}z\right)  ^{2\left(  k-1\right)
^{2}}\nonumber\\
p_{z}  &  =0 \label{22}%
\end{align}
It can be checked that the weak energy condition (WEC) which includes also the
null energy condition (NEC) is not satisfied. WEC says that $\rho\geq0$, and
$\rho+p_{i}\geq0$, for $i=x,y,z$. Since we have $\rho<0$, WEC is violated. It
is observed that for $k\gg1$, $\rho\rightarrow0$ and under that condition WEC
becomes valid.

We conclude the energy discussion by showing simply the limit of
$z_{0}\rightarrow0$. In this limit, we have $\cosh a_{0}z_{0}\rightarrow1$,
and the energy density can be expressed by%
\begin{align}
\rho &  =\lim_{z_{0}\rightarrow0}\left(  -a_{0}^{2}z_{0}\right)  \left(
\frac{\Theta\left(  z\right)  -\Theta\left(  z-z_{0}\right)  }{z_{0}}\right)
\nonumber\\
&  =-\rho_{0}\delta\left(  z\right)  \label{23}%
\end{align}
where $\delta\left(  z\right)  $ is the Dirac delta function.

Let us note that in this limit, we assume that as $z_{0}\rightarrow0$, $\rho_{0}=a_{0}^{2}z_{0}$
remains finite. Even in this limit, we have%
\begin{align}
\rho &  =-\rho_{0}\delta\left(  z\right) \nonumber\\
p_{x}  &  =\rho_{0}\left(  k-1\right)  ^{2}\delta\left(  z\right) \nonumber\\
p_{y}  &  =\rho_{0}k^{2}\delta\left(  z\right) \nonumber\\
p_{z}  &  =0 \label{24}%
\end{align}
which violates the energy conditions.

\subsection{The extrinsic curvature analysis}

In this section, we show that the extrinsic curvature tensor $K_{ij}$ which characterizes the embedding in one higher dimension is
continuous at $z=z_{0}$ \cite{7}. Our metric (\ref{1}) is expressed on the surface
$\sum=z-z_{0}$ by%
\begin{equation}
ds^{2}=d\tau^{2}-f^{2k}dx^{2}-f^{2\left(  1-k\right)  }dy^{2}\label{25}%
\end{equation}
where $d\tau=f^{k\left(  k-1\right)  }dt$ and the $z-$term is dropped out for
$z=z_{0}=$constant. The normal unit vector to $\sum$ is given by%
\begin{equation}
n_{\gamma}=\left(  n_{t},n_{x},n_{y},n_{z}\right)  \label{26}%
\end{equation}
satisfying the condition $n_{\gamma}n^{\gamma}=-1$. The components are%
\begin{align}
n_{t} &  =\frac{1}{\sqrt{\Delta}}\frac{\partial\sum}{\partial t}=0\nonumber\\
n_{x} &  =\frac{1}{\sqrt{\Delta}}\frac{\partial\sum}{\partial x}=0\nonumber\\
n_{y} &  =\frac{1}{\sqrt{\Delta}}\frac{\partial\sum}{\partial y}=0\nonumber\\
n_{z} &  =\frac{1}{\sqrt{\Delta}}\frac{\partial\sum}{\partial z}=\frac
{1}{\sqrt{\Delta}}\label{27}%
\end{align}
where $\Delta=f^{2k\left(  k-1\right)  }$. The components of $K_{ij}$ are%
\begin{align}
K_{\tau\tau} &  =-n_{z}\Gamma_{tt}^{z}\overset{\cdot}{t}^{2}\nonumber\\
K_{xx} &  =-n_{z}\Gamma_{xx}^{z}\nonumber\\
K_{yy} &  =-n_{z}\Gamma_{yy}^{z}\label{28}%
\end{align}
where $\overset{\cdot}{t}=\frac{dt}{d\tau}$ and the Christoffel symbols are%
\begin{align}
\Gamma_{tt}^{z} &  =k\left(  k-1\right)  \frac{f^{\prime}}{f}\nonumber\\
\Gamma_{yy}^{z} &  =\left(  k-1\right)  f^{2\left(  1-k^{2}\right)  }%
\frac{f^{\prime}}{f}\nonumber\\
\Gamma_{xx}^{z} &  =-kf^{2k\left(  2-k\right)  }\frac{f^{\prime}}{f}\label{29}%
\end{align}
Upon substitution of $f\left(  z\right)  =z$ for $z>z_{0}$ and $f\left(
z\right)  =\cosh a_{0}z$ for $z<z_{0}$ we observe that
\begin{equation}
\left(  K_{\tau}^{\tau},K_{x}^{x},K_{y}^{y}\right)  =\left(  -k\left(
k-1\right)  ,-k,k-1\right)  f^{k-k^{2}}\frac{f^{\prime}}{f}\label{30}%
\end{equation}
This leads to the condition, by virtue of $a=0,b=1$ that%
\begin{equation}
\left[  K_{i}^{j}\right]  _{out}-\left[  K_{i}^{j}\right]  _{in}=0\label{31}%
\end{equation}
That is, $K_{ij}$ is continuous at the surface $z=z_{0}$, leading to no source
accumulation on $z=z_{0}$.

\section{Geodesic Analysis of the fall}

In this section, we study geodesic equations of time-like particles for the line-element (\ref{1}). Lagrangian of a free particle is $\mathcal{L}=\frac
{1}{2}g_{\mu\nu}\dot{x}^{\mu}\dot{x}^{\nu}$, which explicitly reads%
\begin{equation}
\mathcal{L}=\frac{1}{2}\left[  f\left(  z\right)  ^{2k\left(  k-1\right)
}\left(  \overset{\cdot}{t}^{2}-\overset{\cdot}{z}^{2}\right)  -f\left(
z\right)  ^{2k}\overset{\cdot}{x}^{2}-f\left(  z\right)  ^{2-2k}\overset
{\cdot}{y}^{2}\right]  \label{32}%
\end{equation}
in which a dot denotes the derivative with respect to proper time, $\tau$. The metric is
independent of $t$, $x$ and $y$, therefore we have three conserved quantities,
energy, linear momenta in $x$ and $y$ direction,which are shown by $E$,
$\alpha_{0}$ and $\beta_{0}$. From the Euler-Lagrange equations we obtain the first integrals from the equations of motion as follows%
\begin{align}
\overset{\cdot}{t}  & =Ef\left(  z\right)  ^{-2k\left(  k-1\right)
}\nonumber\\
\overset{\cdot}{x}  & =\alpha_{0}f\left(  z\right)  ^{-2k}\nonumber\\
\overset{\cdot}{y}  & =\beta_{0}f\left(  z\right)  ^{-\left(  2-2k\right)
}\label{33}%
\end{align}
By the definition of $4-$velocity normalization for massive particle,
$g_{\mu\nu}\frac{dx^{\mu}}{d\tau}\frac{dx^{\nu}}{d\tau}=1$, and replacing
conserved quantities in this equation we get the geodesics equation for
$z$ coordinate as follows%
\begin{equation}
\overset{\cdot}{z}^{2}=f\left(  z\right)  ^{-4k\left(  k-1\right)  }%
E^{2}-\alpha_{0}^{2}f\left(  z\right)  ^{-2k^{2}}-\beta_{0}^{2}f\left(
z\right)  ^{-2\left(  k-1\right)  ^{2}}-f\left(  z\right)  ^{-2k\left(
k-1\right)  }\label{34}%
\end{equation}
Note that for calculating $z\left(  t\right)  $ we  used $\left(  \frac{\overset
{\cdot}{z}}{\overset{\cdot}{t}}\right)  ^{2}=\left(  \frac{dz}{dt}\right)
^{2}$. The result will be%
\begin{equation}
\left(  \frac{dz}{dt}\right)  ^{2}=1-\frac{\alpha_{0}^{2}}{E^{2}}f\left(
z\right)  ^{2k\left(  k-2\right)  }-\frac{\beta_{0}^{2}}{E^{2}}f\left(
z\right)  ^{2\left(  k^{2}-1\right)  }-\frac{f\left(  z\right)  ^{2k\left(
k-1\right)  }}{E^{2}}\label{35}%
\end{equation}
After the second differentiation, we obtain the acceleration%
\begin{equation}
\frac{d^{2}z}{dt^{2}}=-f^{\prime}\left(  z\right)  \left[  \frac{\alpha
_{0}^{2}k\left(  k-2\right)  }{E^{2}}f\left(  z\right)  ^{2k\left(
k-2\right)  -1}+\frac{\beta_{0}^{2}\left(  k^{2}-1\right)  }{E^{2}}f\left(
z\right)  ^{2\left(  k^{2}-1\right)  -1}+\frac{k\left(  k-1\right)  }{E^{2}%
}f\left(  z\right)  ^{2k\left(  k-1\right)  -1}\right]  \label{36}%
\end{equation}
Now we will solve this differential equation for the "outside" of an infinite plate by choosing specific constraints on $k$, to study the behavior of the particle. From now on we make the choice $k\left(  k-1\right)  =\frac{1}{2}$,$\ $
For the "outside" of the infinite plate, we keep the general form $f\left(  z\right)=a+bz$, so that%
\begin{equation}
\frac{d^{2}z}{dt^{2}}=-\frac{b}{2E^{2}}\left[  \alpha_{0}^{2}\left(
1-2k\right)  \left(  a+bz\right)  ^{-2k}+\beta_{0}^{2}\left(  2k-1\right)
\left(  a+bz\right)  ^{2k-2}+1\right]  \label{37}%
\end{equation}
We reexpress Eq (\ref{35}) in the following form in order to derive the speed in terms of energy%
\begin{equation}
1-\left(  \frac{dz}{dt}\right)  ^{2}=\frac{f\left(  z\right)  }{E^{2}}\left[
1+\alpha_{0}^{2}f\left(  z\right)  ^{-2k}+\beta_{0}^{2}f\left(  z\right)
^{2k-2}\right]  \equiv\frac{f\left(  z\right)  G\left(  z\right)  }{E^{2}%
}\label{38}%
\end{equation}
where%
\begin{equation}
G\left(  z\right)  \equiv1+\alpha_{0}^{2}f\left(  z\right)  ^{-2k}+\beta
_{0}^{2}f\left(  z\right)  ^{2k-2}\label{39}%
\end{equation}
and by imposing the initial values of particle, $\nu_{i}=$ the initial speed, $f_{i\text{ }}$, $G_{i}$, at $t=t_{i}$ we have%
\begin{equation}
\frac{1}{E^{2}}=\frac{1-\nu_{i}^{2}}{f_{i\text{ }}G_{i}}\label{40}%
\end{equation}
It is seen from (\ref{36}) that for the motion perpendicular to the plate
$\left(  \alpha_{0}=0=\beta_{0}\right)  $ we obtain%
\begin{equation}
\frac{d^{2}z}{dt^{2}}=-\frac{b}{2E^{2}}=-\left(  \text{positive constant}%
\right)  \label{41}%
\end{equation}
In particular, the free fall problem $\left(  \nu_{i}=0\right)  $, gives%
\begin{equation}
z=z_{i}-\frac{b}{4E^{2}}\left(  t-t_{i}\right)  ^{2}\label{42}%
\end{equation}
as in the Newtonian fall. Near to the plate $z\thickapprox z_{0}$ the
acceleration turns out to be $a=\frac{b}{2E^{2}}$. Which determines the G-force of the plate in acceleration with Newoton's law.

For comparable momenta $\alpha_{0}$ and $\beta_{0}$, we observe that the acceleration (\ref{36}) remains negative. Far away from the plate $\left( 
z\gg z_{0}\right)  $ attraction holds also true irrespective of $\alpha_{0}$
and $\beta_{0}$. The ambiguity of the repulsive effect, however, can be removed completely by choosing $\alpha_{0}>\beta_{0}$ for $2k-1=-\sqrt{3}\left(
\text{ and }\beta_{0}>\alpha_{0}\text{ for }2k-1=+\sqrt{3}\right)  $, which are the admissible values for $k$ in the assumed constraint $k\left(  k-1\right)  =\frac{1}{2}$.

\section{Conclusion}

We considered the class of Kasner metrics depending on the space
coordinate $z$ instead of time as a representation of a plate of
thickness $z_{0}$. We have studied the possibility of falling on such a plate.
In Newtonian gravity near the Earth, an object makes a free fall motion with a
constant negative acceleration. In the presence of repulsive forces of general
relativity metrics, we remind that these requirements of fall are not
automatically satisfied. Our plate metric fails to satisfy the weak energy
condition (WEC) whose violation becomes minimized for the free parameter $k\gg1$ or, as in other physical systems in order to provide physicality of our plate metric other physical fields must be coupled.
On the other hand, the plate metric is entirely regular whose Kretchmann
scalar is finite everywhere. It is observed that the choice of the stress tensor
$T_{zz}=0$, is effective in making everything regular. The geodesic analysis
shows that for certain $k$ values falling with a constant negative
acceleration holds true as in Newton's theory. Specifically, $k=\frac{1{\mp}\sqrt{3}}{2}$, proves to be
a good choice which corresponds to the Kasner exponents $p_{1}=\frac{1}{3}$,
$p_{2}=\frac{1{\mp}\sqrt{3}}{3}$ and $p_{3}=\frac{1{\pm}\sqrt{3}}{3}$. The tuning of
momenta along $x$ and $y$ makes also the fall but the acceleration is
not a constant. Finally, we comment that a comparison of our plate geometry with
the domain wall spacetime \cite{2}\cite{3}\cite{4} suggests that the thin-wall boundary can be
extended to a thick-wall in a special direction.

\end{document}